\documentclass[aps,pre,reprint,amsmath,amssymb, superscriptaddress,doublecolumn]{
revtex4-1}
\usepackage{graphicx}
\usepackage{epsf}
\usepackage{dcolumn}
\usepackage{bm}
\usepackage{hyperref}
\usepackage{latexsym}
\usepackage{color}
\usepackage[english]{babel}
\usepackage{amssymb}
\usepackage{amsmath}
\usepackage{natbib}
\usepackage{xcolor}

\usepackage{caption}
\usepackage{subcaption}

\newcommand{\beq}{\begin{equation}}
\newcommand{\eeq}{\end{equation}}

\newcommand{\bsy}{\boldsymbol}

\newcommand{\pbar}{\bar{p}}
\newcommand{\qbar}{\bar{q}}
\newcommand{\Ncal}{\mathcal{N}}

\begin{document}
\def\av#1{\langle#1\rangle}
\def\etal{{\it et al\/.}}
\def\pc{p_{\rm c}}
\def\l{{\lambda}}
\def\hm{h_*}
\def\xm{x_*}
\def\remark#1{{\bf *** #1 ***}}
\def\ket#1{|#1\rangle}
\def\bra#1{\langle#1|}
\def\braket#1#2{\langle#1|#2\rangle}

\title{Robust secret storage in networks}

\author{Vinko Zlati\'{c} }

\affiliation{Theoretical Physics Division, Rudjer Bo\v{s}kovi\'{c} 
Institute, P.O.Box 180, HR-10002 Zagreb, Croatia}

\begin{abstract} The problem of storing secure information on a network is studied. A formal framework for distributed secret storage is introduced, and possible applications in technological and social systems are discussed. The problem is formulated as the optimization of a robustness functional in which two competing requirements are balanced: survivability under network-degrading processes and resistance to adversarial compromise. An exact representation of survivability is derived in terms of minimal information-carrying subgraphs (MICS), which provide a reduced description of the reconstruction events relevant to the stored information. This representation is then used to construct semi-local optimization methods whose dynamics do not require global knowledge of the network structure. Finally, it is shown that, in a limiting case, the robustness functional can be mapped naturally to an effective spin Hamiltonian.

\end{abstract}
\maketitle

\section{Introduction}
Information security is most commonly associated with encryption~\cite{aumasson2017serious}, but another fundamental strategy is to split information into several pieces and distribute them among different agents. This idea is formalized in secret sharing, where a secret is partitioned into $N$ shares and can be reconstructed only when a prescribed subset, or at least $m<N$ shares, is available~\cite{slinko2020algebra,cramer2015secure,shamir1979share}. While the cryptographic construction of the shares has been extensively studied~\cite{aumasson2017serious,herzberg1995proactive}, much less attention has been given to the network question of where and how these shares should be stored. In networked systems,
however, the placement of shares is not a secondary issue: the probability of
recovering the secret depends not only on the cryptographic scheme but also on
which vertices store which pieces and how failures or compromises propagate
through the underlying graph.

Related questions have appeared in models of secret reconstruction under network failures~\cite{lee1999probability}, graph-theoretic placement of secrets~\cite{poguntke2016near}, confidentiality--availability trade-offs under cyber threats~\cite{zhang2021confidentiality}, and distributed replica placement~\cite{kohler2015self}. These problems are also connected to classical network-reliability and resource-placement problems, which are computationally difficult in general~\cite{johnson1994complexity}. The present work differs by formulating a single robustness functional that explicitly trades survivability under failures against hackability under compromises, and by asking when this functional can be approximated and optimized from local network information.

This question also belongs to the broader physics of network resilience, where failures, percolation, and the stability of dynamical processes have been studied extensively~\cite{callaway2000network,derenyi2005clique,paul2006optimization,radicchi2009explosive,serrano2011percolation,karrer2014percolation,allard2019percolation}. However, the resilience of secure distributed information remains comparatively less developed. Previous work has considered small graphs optimized to preserve information under failure or hacking~\cite{vojkovic2018multicoloring,vojkovic2023graph}, and the transmission of secrets through networks with shared vulnerabilities led to color-avoiding percolation~\cite{krause2016hidden,krause2017color,kadovic2018bond}.

Here we study a complementary storage problem: information is first partitioned by an available cryptographic or secret-sharing scheme, and the central question is how the resulting symbols should be placed on the vertices of a network. We show that the exact survivability is governed by minimal information-carrying subgraphs, use them to construct semi-local robustness functionals, and test when these functionals provide useful surrogates for optimization.

One envisioned application and main motivation for this work is future trusted torrent-like storage
systems. The benefit of not storing the whole information on a local computer is
twofold: first, it provides an additional level of protection and privacy; and
second, it reduces the amount of local memory needed, because large parts of the
information are stored on other people's computers. A detailed discussion of the
implementation of such systems is beyond the scope of this paper.

\section{Problem formulation}
A given network (graph) $\Gamma$ is present. The adjacency matrix $\bsy{\hat{A}}$ completely describes this graph.
An \emph{N-separable information} defined as  $\bsy{X}\equiv\{X_{\mu}\}\equiv\{X_1,X_2,\ldots X_N\}$ is also present. N-separable information refers to an array of symbols (bits) that can be partitioned into N pieces. It is assumed that the information is completely useless if not all $N$ pieces of information are obtained together. A more general case of $m$ pieces of information needed to reconstruct information, separated in $N$-pieces would be of additional practical interest but is beyond the scope of this paper.


The goal is to store the information in the network robustly, optimizing for the following conditions:\emph{(i)}The network should be as resilient as possible to the failure i.e. the disappearance of some vertex in the network should not eradicate a connected component which still contains the whole information. \emph{(ii)}The adversary (hacker) should have a problem collecting the information—he has to hack as much as possible of the information-preserving vertices.
 
It is clear that problem presented above is too broadly defined, as there are many different formulations one could use with respect to real-world applications. Furthermore, depending on the application, different additional constraints could be imposed, such as the size of the memory, the possibility that hackers can more easily access better-connected vertices, etc.
In the following analysis, the focus is on the simplest instance of the problem that also takes into account the particular topology of the network structure in which the information is stored. 

\begin{figure}[t]
  
\includegraphics[width=0.45\textwidth]{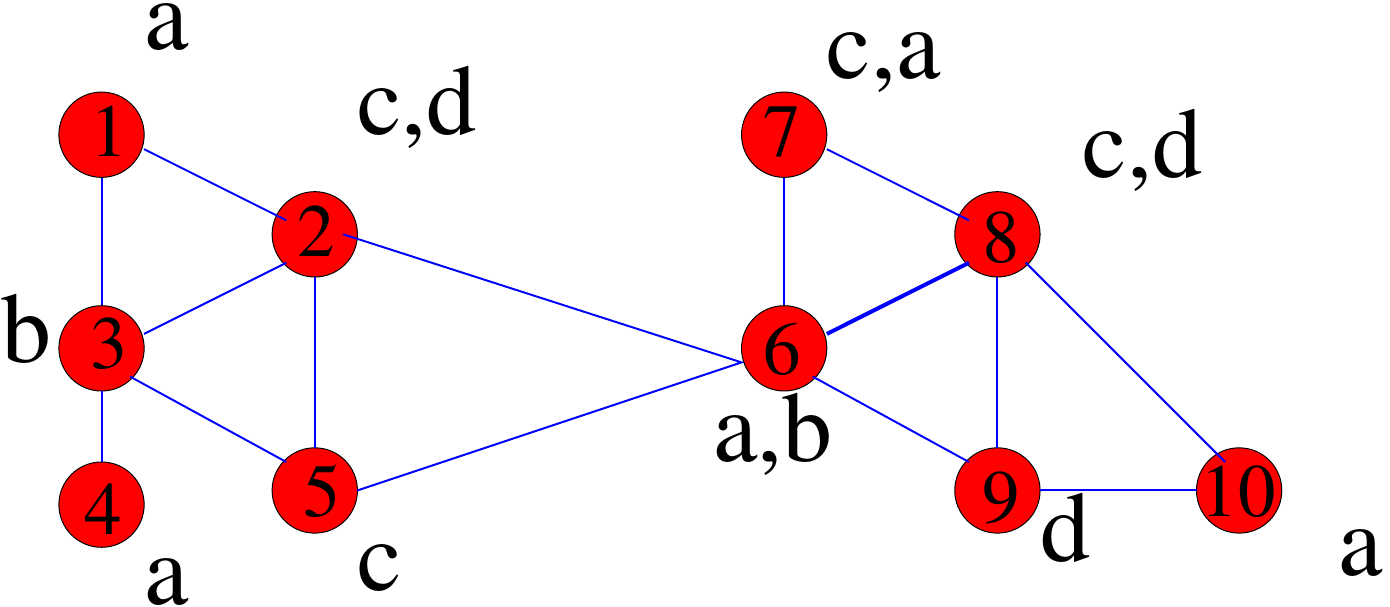}
{\Large a)}
\includegraphics[width=0.45\textwidth]{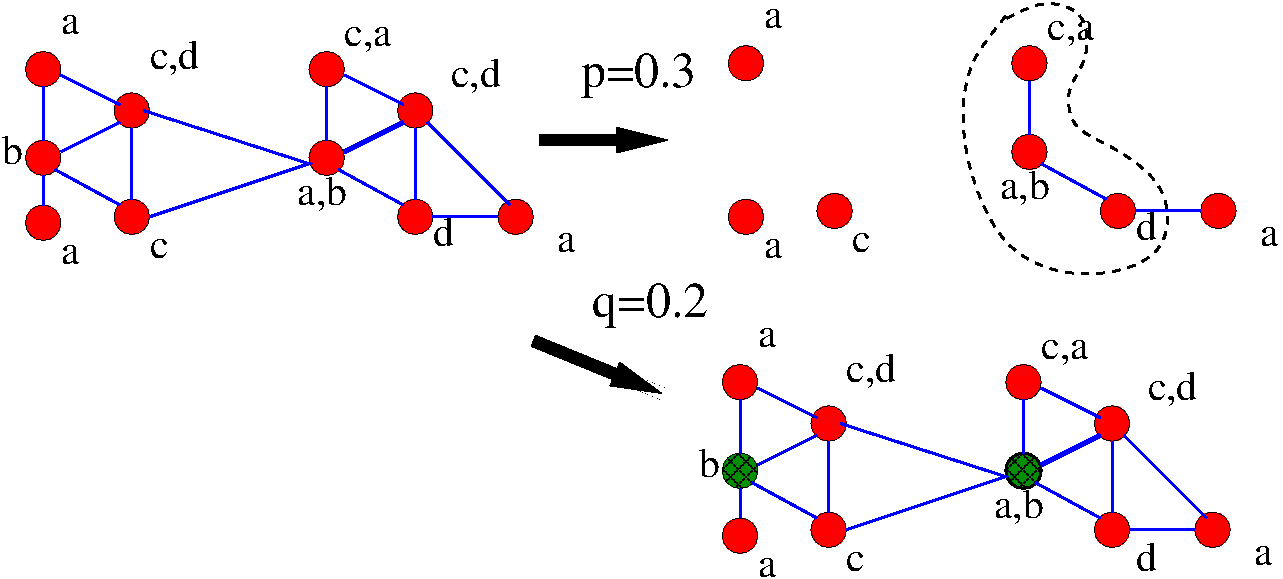}
{\Large b)}

\caption{This figure represents one network of 10 vertices with one possible configuration of
information given by the alphabet of 4 symbols (a,b,c,d). In the panel a) at the top is shown the
original configuration of information in the network.
In the lower panel b), the upper right side represents the network
which lost 3 of its vertices due to network failure. Although the network is severely
damaged the complete information has survived in the designated part of the network. The
lower right side represents the vertices that were efficiently hacked by hackers who had
0.2 probability of hacking any vertex.  }
\label{fig:Spread}
\end{figure}

The first condition traditionally translates to the proposition that every vertex has the same
probability $p$ to be deleted from the network due to some network failure process.
Similarly, the second condition can be translated to the proposition that a hacker has a certain probability $q$ to collect (hack), a piece of information from some vertex. 
 A possible evolution in this system is shown in
Fig.\ref{fig:Spread}.
%

The problem is reduced to its simplest instance by this interpretation. Models that better describe reality (such as changing the probability of hacking for a vertex if its neighbor is hacked, increasing the probability of vertex deletion based on the number of deleted neighbors, considering the spatial embedding of vertices where space coordinates influence the assessment of probabilities $p$ and $q$, etc.) can easily be expanded. Furthermore, for some structures, like for example trees, one might consider optimizing for the number of components that still contain the whole information. 
The solving of this problem is a first step towards addressing more complex and realistic security problems.


Quantities used to measure the optimization of this problem shall be defined hereafter. Pieces of information will be named $symbols$. The \emph{configuration} $\chi(\bsy{X},\Gamma)$ of the information on the network is defined as a mapping between sets of symbols and vertices of the graph, i.e., $\chi:V\rightarrow \Omega(\bsy{X})$, where $V$ represents the set of all vertices of the graph $\Gamma$, and $\Omega(\bsy{X})$ represents the set of all possible distinct collections of information symbols $X_{\mu}$ i.e.
$\Omega(\bsy{X})=\{X_1,X_2,...,X_N,X_1X_2,...,X_{N-1}X_N,...,X_1X_2...X_N\}$, except for the empty set $\varnothing$, since the assumption is that each vertex contains at least one symbol. For example, the configuration of the most left lower vertex in Fig.\ref{fig:Spread} corresponds to the symbol $\{a\}$, while the configuration of the upper right vertex in all three figures corresponds to the symbols $\{a,c\}$. 


The \emph{survivability} $\mathcal{S}(p,\chi)$ is defined as a probability that the information is completely preserved in at least one surviving subgraph of the original
network after the network crash described by the parameter (probability) $p$ if the initial configuration of information is given with the configuration $\chi$. 

The \emph{hackability}
$\mathcal{H}(q,\chi)$ is defined as the probability that the hacker will obtain complete
information from hacked vertices after the process of hacking described with parameter $q$
is finished if the configuration was $\chi$.

Clearly one wishes to increase survivability as much as possible while reducing the hackability for some reasonable choice of parameters $p$ and $q$. Here, the assumption is made that parameters $p$ and $q$ can be inferred from the historical statistics or at least estimated from the more detailed knowledge of the system's robustness and security. Throughout the text, $\bar{p}=1-p$ is also used for the probability that a vertex survives the deletion process. In Appendix~\ref{appA}, a set of illustrative examples is presented to complement the main discussion. These examples are intended to make the preceding arguments more concrete, and intuitive.

In order to find robust configurations for different networks, there is a need to connect
survivability and hackability of the network into one unique functional. The easiest way to do this is to add one more parameter $\alpha$. The parameter $\alpha$ is defined as the probability that a network will be exposed to a process of deletion if some event occurs, and $(1-\alpha)$ is the probability that the network will be attacked by the hacking process if some event occurs. Then the most natural functional to be optimized is
\begin{equation}\label{FirstFunctional}
 \mathcal{F}(\alpha,\mathcal{S},\mathcal{H})=\alpha\mathcal{S}+(1-\alpha)(1-\mathcal{H}).
\end{equation}

It is just the probability that the information will be preserved and secured if some event happens. Written in a more formal way it is the conditional probability that the information has survived an event given the parameters, network structure, and configuration $\chi$ i.e. 
\begin{equation}\label{FunctionalCOndProb}
 \mathcal{F}(\alpha,\mathcal{S},\mathcal{H})=P(I|\alpha,p,q,A,\chi),
\end{equation}

in which $I$ designates an event in which information survived and was not hacked. It is again assumed that parameter $\alpha$ can be inferred from historical data on the number of hacking and network failure events. It will immediately be clear that the computation of exact survivability and hackability functionals is not easy in the most general case. I


Survivability can be defined in the following way. Find all
distinct subgraphs $\Gamma'$ which result after the process of network destruction, then assign to them the probability of survival which is just
$\mathcal{P}(\Gamma')=(1-p)^{|\Gamma'|}p^{|\Gamma|-|\Gamma'|}$. Then survivability is defined as
\beq
\mathcal{S}(p,\chi,\Gamma)=\sum_{\Gamma'}\mathcal{P}(\Gamma')\delta_{\bsy{X}\Gamma'},
\label{eq:Survivability-exact}
\eeq
where $\delta_{\bsy{X}\Gamma'}=1$ if there is at least one connected subgraph of $\Gamma'$
which contains the whole information $\bsy{X}$ and $0$ otherwise. For sufficiently large
networks it is clearly hard to compute all of the resulting $\Gamma'$ and it is
even harder to check if the information is fully preserved in some of the associated
connected subgraphs of $\Gamma'$.

Hackability can be defined in the same manner. Divide the vertices of the network in
two sets in all possible ways - one set representing hacked vertices and other representing vertices that were not hacked. Associated probability of every division $D$ is
$\mathcal{P}(D)=q^{|D_h|}(1-q)^{(|\Gamma|-|D_h|)}$, where $|D_h|$ is the size of the set
of vertices that got hacked.
The hackability is then clearly
\beq\label{eq:HackabilityExact}
\mathcal{H}(q,\chi)=\sum_{D}\mathcal{P}(D)\delta_{\bsy{X}D_h},
\eeq
Where $\delta_{\bsy{X}D_h}=1$ if all information is contained in the hacked set. In the following text $1-\mathcal{H}\equiv{\bar{\mathcal{H}}}$ is often used.

\section{Approximation of Robustness functional}

For large networks of interest, the exact calculations will not be possible. Therefore, it
is vital for the possible applications to devise intelligent approximation schemes for
realistic problems. However, for smaller networks, there is a way to compute intelligently $\mathcal{S}$ exactly.

First, the equation \ref{eq:Survivability-exact}, can be rewritten taking into account component sizes.  Notice that equation \ref{eq:Survivability-exact} can be interpreted as a product of two independent probabilities, the first one computing the probability that the subgraph $\Gamma'$ has survived the process of network destruction and the second one $\delta_{\bsy{X}\Gamma'}$ that is equal to 1 if event \emph{Information exist} happened and 0 if it did not. As long as in a given configuration, there is no complete information present in the monitored subgraph, the contribution $\mathcal{S}$ is zero. Clearly, if written only in terms of $\bar{p}$, $\mathcal{S}$ will be a polynomial. In appendix \ref{app:derivations}, it is shown that  $\bar{\mathcal{S}}$ can be written as a polynomial in the following way:

\begin{equation} \mathcal{S}(p,\chi,\Gamma) = \sum_{r=1}^{|\Gamma|} a_r(\chi,\Gamma)\,\bar p^r , \label{eq:SurvivabilityEXACT} \end{equation} where the coefficient of $\bar p^r$ is \begin{equation} a_r(\chi,\Gamma) = \sum_{\ell=1}^{M} (-1)^{\ell+1} \sum_{\substack{ 1\leq j_1<j_2<\cdots<j_\ell\leq M\\ |\tilde{\Gamma}'_{j_1}\cup\tilde{\Gamma}'_{j_2}\cup\cdots\cup \tilde{\Gamma}'_{j_\ell}|=r }} 1 . \label{eq:SurvivabilityCoefficient} \end{equation}

Here $\tilde{\Gamma}'$ represent  \emph{minimal
information-carrying subgraph} (MICS), i.e. the connected subgraph that contains the whole information and that does not have proper subgraphs that also contain the whole information.

In the table \ref{Tablica2}, the MICS of the graph presented in panel a) of Fig. \ref{fig:Spread} that are contributing to Survivability defined in \ref{eq:SurvivabilityEXACT}, are outlined as an example for the reader. 

\begin{table}
\centering
\begin{tabular}{|l|l|}
\hline 
$|\tilde{\Gamma}'|$ & $\tilde{\Gamma}'$  \\
\hline
\hline
2 & $\{2,6\}$ \\
2 & $\{6,8\}$ \\
\hline
3 & $\{1,2,3\}$   \\
3 & $\{2,3,4\}$  \\
3 & $\{5,6,9\}$  \\
3 & $\{6,7,9\}$   \\
\hline
\end{tabular}
\caption{MICS of the graph $\Gamma$  and configuration $\chi$ presented in fig. \ref{fig:Spread} that contain complete information. The exact Survivability is in this case $S=2\pbar^2+3\pbar^3-8\pbar^4+5\pbar^5-\pbar^6$} 
\label{Tablica2}
\end{table}

In the appendix \ref{app:derivations}, it is shown that $\bar{\mathcal{H}}$ can be written as:

\begin{align}
    \bar{\mathcal{H}}&=\sum_{i}\bar{q}^{\mathcal{N}(X_i)}-\sum_{i< j}\bar{q}^{\mathcal{N}(X_i)\circ \mathcal{N}(X_j)}+\nonumber\\
    &+\sum_{i< j< k}\bar{q}^{\mathcal{N}(X_i)\circ \mathcal{N}(X_j)\circ \mathcal{N}(X_k)}+\ldots.
    \label{eq: H_EXACT_Summation}
\end{align}

Target functional $\mathcal{F}$ can then be exactly computed using the equations \ref{eq:SurvivabilityEXACT}, \ref{eq: H_EXACT_Summation} and \ref{FirstFunctional}. Optimizing this functional for larger networks can be an unattainable task. 

One thing that can be done is to write $\mathcal{F}$ as an approximation, using the fact that in the general case of vertex failure, the size of surviving subgraphs will decay exponentially with the size of subgraphs~\cite{newman2001random,kryven2017general}. This means that the majority of surviving components will be small, and one can approximate the equation \ref{eq:SurvivabilityEXACT} by limiting the maximal size of subgraphs $|\Gamma_M|$ that will be used for computation, which directly affects the maximal order of the polynomial that will be used for computation.  The whole structure of graph $\Gamma$ is not needed but only the local contribution from the neighborhood of the vertex $i$, whose radius $R$ is determined by the $|\Gamma_M|$. In this way the computation of optimal robustness $\mathcal{F}$ does not involve complete information of the structure although the state depends on the totality of the graph $\Gamma$. 
The semi-local survivability is then approximated by 
\begin{equation} \mathcal S_{R1}(p,\chi,\Gamma) = 1-\prod_{v\in V}\left(1-P_v^{(R)}\right). \label{eq:semi-local-survivability1} \end{equation}

In the case of the smaller graph one can use more refined approximation to survivability:
\begin{equation} \mathcal S_{R2}(p,\chi,\Gamma) = \sum_{\emptyset\neq \mathcal C\subseteq V} \; \sum_{\substack{ (\mathcal A_v)_{v\in\mathcal C}\\  \mathcal A_v\subseteq \widetilde{\mathcal G}_v^{(R)}  }} (-1)^{1+\sum_{v\in\mathcal C}|\mathcal A_v|} \, \bar p^{ \left| \bigcup_{v\in\mathcal C} \Gamma'_{\mathcal A_v} \right| }, \label{eq:semi-local-survivability2} \end{equation}

 where $\mathcal A_v$ is the nonempty subfamily of the local family $\widetilde{\mathcal G}_v^{(R)}$ of MICS rooted in $v$, and the equation is just the inclusion-exclusion equation that omits MICS larger than the neighborhood.  
 The details of derivation are presented in App. \ref{app:semi-local-functional}. 
 Furthermore, Hackability \ref{eq: H_EXACT_Summation} is also a polynomial that can be approximated, taking into account the order of approximation applied to Survivability. The approximation is most convenient for the computation of optimal configuration, as it can be performed locally on the vertex $i$ that is changing the state in, for example, simulated annealing.
 
 Estimated error from such a local approximated functional is :  
\begin{equation}
|\mathcal F-\mathcal F_R|
\leq
\alpha\epsilon_{\rm cover}+ (1-\alpha)\epsilon_{\rm hack},
\label{eq:functional-error-bound}
\end{equation}
where for a locally tree-like graph with average degree $z$, the rough estimate
$N_m\lesssim |\Gamma|z^{m-1}$ gives
\begin{equation}
\epsilon_{\rm cover}=\mathcal S-\mathcal S_M
\lesssim
|\Gamma|\,\pbar\,
\frac{(z\pbar)^M}{1-z\pbar},
\qquad z\pbar<1 .
\label{eq:tree-tail-error}
\end{equation} The term $\epsilon_{\rm hack}$ is estimated through Bonferoni bounds 
\begin{equation}
B_K=
\sum_{1\leq |T|\leq K}
(-1)^{|T|+1}\qbar^{c_T},
\end{equation}
and is 
\begin{equation}
\epsilon_{\rm hack}=\left|\bar{\mathcal H}
- \bar{\mathcal H}_{2m}^{\rm mid}
\right|
\leq
\frac{B_{2m-1}-B_{2m}}{2}.
\end{equation}

The details of the error derivation are in App:~\ref{app:semi-local-message-passing}.

\section{Tests}

\begin{figure}[t]
 \includegraphics[width=0.5\textwidth]{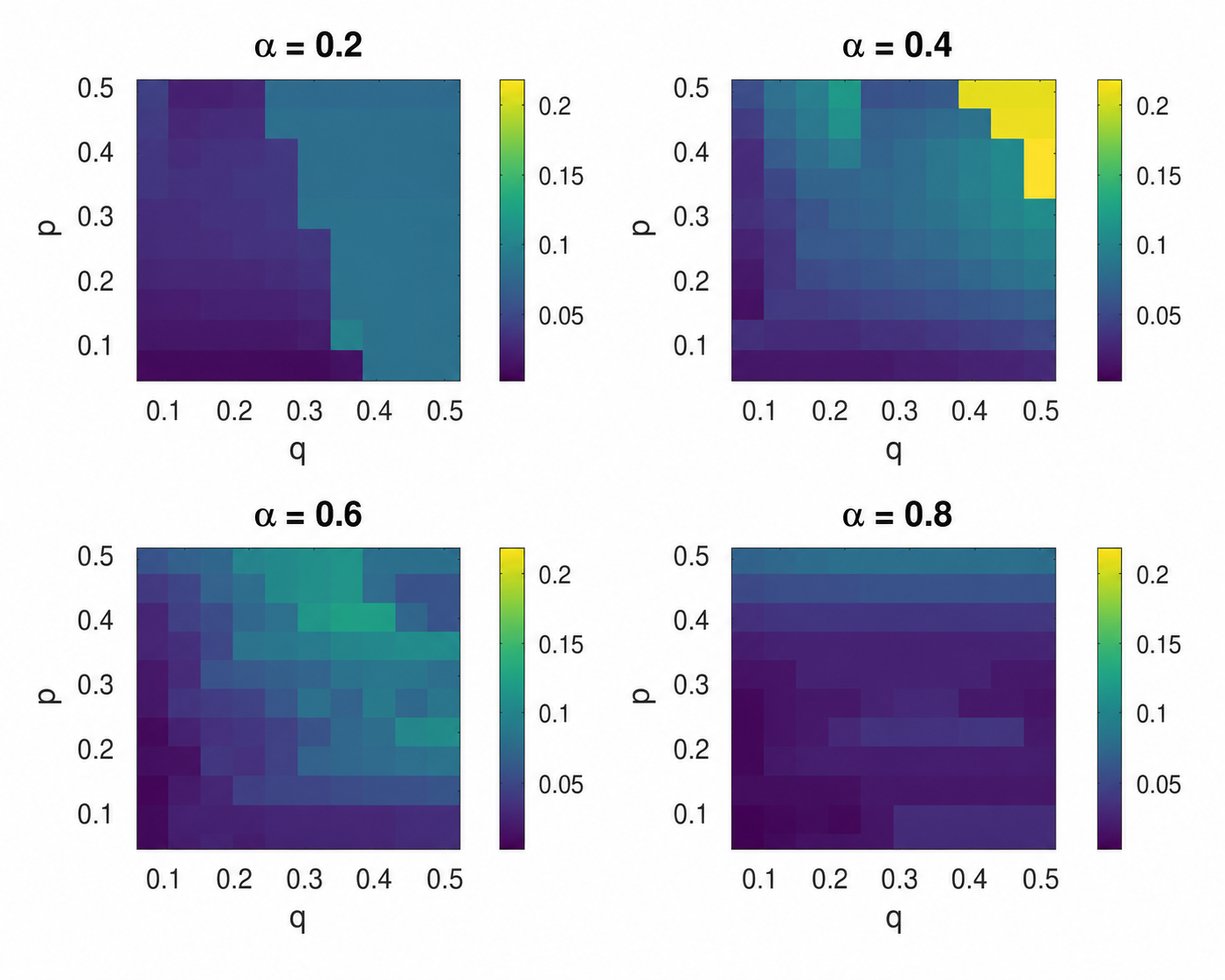}
\caption{ The plot of $(\mathcal{F}-\mathcal{F}_{app})/\mathcal{F}$ for $\alpha\in\{0.2,0.4,0.6,0.8\}$ and different values of $p$ and $q$. The network is ER of size $N=20$ and $z=4$. The number of symbols is $3$. For larger networks, the computation of the true functional is not feasible. }
\label{fig:Differences}
\end{figure}

To test the validity of the presented approach, the number of smaller networks $N=8,10,12,14,16$ for which value of the functional given configuration could be computed exactly is compared to two approximations in Table~\ref{tab:semilocal-validation-summary}.  The R2 approximation is consistently much more accurate than R1, while the residual finite-size fluctuations are visible across network sizes. It is expected that the overlap mistake is becoming smaller. The radius of the boundary is taken to be one.   
 Max-sum message passing was used only as a semi-local heuristic for producing candidate symbol placements~\cite{mezard2009information}. The algorithm is described briefly in App.~\ref{app:semi-local-message-passing}. It performs well for small symbol sets, especially $N_{\rm sym}=3$--$4$, and allows tests on networks larger than those accessible to exact enumeration. 
The approximate computation for the system for which exact robustness function was feasible to compute (N=20 vertices, average degree $z=4$), agrees very well between the approximate and exact computation (see fig.~\ref{fig:Differences}).

For networks with many symbols or large local neighborhoods, the factor tables grow exponentially, so more specialized optimization methods would be required.

\begin{table}[t]
\centering
\caption{Comparison between the exact robustness functional and the semi-local approximations for $4$ symbols on samples of 10 ER networks. Errors are averaged over all tested values of $p\in\{0.2,0.4,0.6\}$, $q\in\{0.05,0.1,0.2\}$, and $\alpha\in\{0.2,0.5,0.8\}$. The $\langle\Delta F_{R1}|\rangle$ and $\langle |\Delta F_{R2}|\rangle$ are absolute means of errors, while $\langle \epsilon_{F,R1}\rangle$ and $\langle \epsilon_{F,R2}\rangle$ are means of relative errors. }
\label{tab:semilocal-validation-summary}
\begin{tabular}{ c c c c c c}
\hline
 $N$ & z & $\langle |\Delta F_{R1}|\rangle$ & $\langle |\Delta F_{R2}|\rangle$ & $\langle \epsilon_{F,R1}\rangle$ & $\langle \epsilon_{F,R2}\rangle$ \\
\hline
 8 & 3.00 & 0.032 & 7.85e-6 & 0.0409 & 9.7e-6 \\
 10 & 3.00 & 0.029 & 1.16e-5 & 0.0117 & 1.47e-5 \\
12 & 3.00 & 0.0189 & 8.6e-4 & 0.0382 & 1.11e-3 \\
 14 & 3.00 & 0.00913 & 4.1e-4 & 0.0121 & 5.36e-4 \\
 16 & 3.00 & 0.0083 & 2.42e-4 & 0.0244 & 3.21e-4 \\
\hline
\end{tabular}
\end{table}

\section{spin systems}

Besides statistical physics reasoning that was used to develop the problem, there is also a more profound connection between this problem and usual spin models.
If the information is separated only into two symbols, the system is easy to compute exactly as there are contributions only from vertices and edges. Any MICS that contains the information and contributes to the functional is either a single vertex that contains the total information of two different symbols or an edge with two different symbols on each vertex. 

For this example, it is easy to verify that the equation for robustness for large $\alpha$ or very small $q$ is:

\begin{align}
 \mathcal{F}=&\alpha\left(1-p^{n_{X_1X_2}}\left(1-\pbar^2\right)^{l_{X_1X_2}}\right)\nonumber\\
 &+(1-\alpha)\left(\qbar^{\Ncal_{X_1}}+\qbar^{\Ncal_{X_2}}-\qbar^N\right).
\end{align}

Here $l_{X_1X_2}$ represents the number of edges that contain the symbol $X_1$ on one of its vertices and the symbol $X_2$ on the other. 

In the case of small values of  $\pbar$ and small values of $q$, one can write the following energy, see App~\ref{app:spin}:

\begin{equation}
\mathcal{F}
\simeq
1-\alpha
+
hn_{X_1X_2}
+
J
\left(l_{X_1X_2}-\binom{n_{X_1X_2}}{2}\right)
\label{eq:spin-one-functional}
\end{equation}
where
\begin{equation}
h=\alpha\pbar-(1-\alpha)q,
\qquad
J=\alpha\pbar^2 .
\end{equation}
Equivalently, defining an effective energy
$\mathcal{E}=-\mathcal{F}$ and dropping the additive constant, we obtain
\begin{equation}
\mathcal{E}
=
-hn_{X_1X_2}
-
J
\left(l_{X_1X_2}-\binom{n_{X_1X_2}}{2}\right).
\label{eq:spin-one-hamiltonian}
\end{equation}

It is easy to see that $h$ can be interpreted as an external field acting on vertices with two symbols, while $J$ is equivalent to anti-fero term for single-symbol vertices and the global repulsive interaction between vertices with two symbols.
The solution of the ground state for a 1D chain shows interesting behavior in which one side of the chain contains a finite cluster of vertices with double symbols, while the rest of the chain consist of alternating one-symbol states (see App.~\ref{app:spin}).

\section{discussion}

A network formulation of robust secret sharing has been presented. Two competing requirements have been formalized: the information should remain recoverable after random failures of vertices, while it should remain difficult to reconstruct from randomly compromised vertices. These requirements have been quantified by the survivability $\mathcal S$, the hackability $\mathcal H$, and the combined robustness functional $\mathcal F$. Exact expressions have been derived in terms of MICS and inclusion-exclusion formulas, making explicit how network topology and symbol placement jointly determine the robustness of the stored information.

Approximate and local formulas for evaluating and optimizing this functional have also been developed. MICS have been used to obtain polynomial representations of survivability, while the hackability contribution has been expressed through symbol-subset counts. These formulations have enabled semi-local approximations, and optimization schemes based on local neighborhoods. The resulting framework shows that the placement of secret symbols can be an important part of secure information storage. This research provides a systematic basis for designing distributed storage systems in which security and robustness are optimized together.

The first thing important to outline is that the presented method can be computed in a distributed setting using only local information, limited to the maximal subgraph size of interest, i.e. a reasonable approximation of the optimal configuration can be achieved even if the structure of the network is not known to any central planner. This property should be useful in different secret sharing networks, like, for example, distributed torrenting. However, it is also important to stress that the proposed methodology works for relatively small sets of symbols. If the size of the symbol set is very large, the proposed method is unfeasible, and more appropriate methods should be developed.  

The proposed framework connects security of information storage to problems related to statistical physics, Using some of the more advanced algorithms developed in the community could greatly influence usefulness and adoption of this framework. 

It was already mentioned that further refinement, like storage capacity, more complex hacking strategies, etc., can be tackled using this framework. Furthermore, although the model is framed for the secret sharing, it is straightforward to see that possible applications of the (perhaps slightly modified) framework could include critical infrastructures and critical product chains, microbial consortia of species exchanging essential metabolites, functional biodiversity, and also possibly healthcare service networks
or any other capability networks.  

\section{Acknowledgments}

The author wants to express gratitude to Sebastian Morel-Balbi and Lucija Nora Farka\v{s}, that have critically read the paper. The author also acknowledges the support
from Croatian science foundation project HRZZ IP-2022-10-1648. This research was also supported by the QuantiXLie Center of Excellence, a project co-financed by the Croatian Government and European Union through the European Regional Development Fund - the Competitiveness and Cohesion Operational Programme (Grant KK.01.1.1.01.0004). 

\bibliography{SecureEmbedingReferences}

\appendix
\section{Illustrative examples for survivability and hackability}
\label{appA}
In order to better understand the problem it is useful to consider how to optimize one condition at the time without considering the other.

What configuration $\chi_s$ will maximize the probability that the information will survive
the failure of a certain portion of the network? The answer is simple, copy \emph{all of the information} on every given vertex i.e. $\chi_s(i)=X_1X_2...X_N \forall i$. In this way even if all the vertices except one get destroyed the information will be preserved i.e. the survivability is given with
$\mathcal{S}(p,\chi_s)=1-p^{|\Gamma|}$, where $|\Gamma|$ is the
size of the graph i.e. the number of vertices. In this limiting case, survivability will not depend on the network structure.

While such a configuration will maximize the survivability of the information, its impact on the hackability will be unsatisfactory. Namely the hackability
$\mathcal{H}(q,\chi_s)$ defined as the probability that a hacker will collect the
whole information by randomly attacking vertices with probability $q$ will be large
i.e. $\mathcal{H}(q,\chi_s)=1-(1-q)^{|\Gamma|}$. This just means that a hacker will obtain the whole information if he or she successfully hacks just one vertex of the network.

Now, which configuration will minimize the hackability of the network? For the case $|\Gamma|\geq N$ The hackability will be minimal if $\chi_h(i)=X_i \forall i\in\{1,N-1\}$ and
$\chi_h(i)=X_N \forall i\in\{N,|\Gamma|\}$ i.e. every symbol except the last is
spread over exactly one vertex in the graph taking care that no vertex possesses more than one
symbol while the last symbol is spread over all the other vertices. Note that this configuration is $N-$fold degenerate. For this choice of configuration hackability will be
$\mathcal{H}(q,\chi_h)=q^{N-1}(1-(1-q)^{|\Gamma|-N+1})$, but the survivability will be at most $\mathcal{S}(p,\chi_h)\leq(1-p)^{N-1}$ because if any of the vertices
which
contain unique symbol is deleted the information will not be preserved. In the case in which the network is a complete graph, this expression is exact.

In these examples, the structure of the network did not really matter. However, it is clear that the structure influences the survivability of the information, while the hackability is not a function of a network structure in this formulation of the problem. Still, as a note it is valuable to mention that in different instances of the problem hackability could also depend on network structure. For example, if in the problem statement is assumed that the hacking probability $q$ of the vertex can be changed if some of its neighbors were previously hacked. Such a modification is definitely realistic and of enormous practical interest, but is beyond the scope of this paper in which the simplest formulation of the problem is considered.
\begin{figure}
 \includegraphics[width=0.48\textwidth]{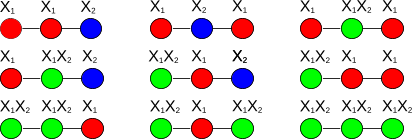}
\caption{ In this figure are presented 9 distinctly different configurations of 2-symbol
information in this simple graph. Associated survivability and hackability are listed in
table \ref{Tablica3vertices}.}
\label{fig:9Spreads}
\end{figure}

\begin{table}
\centering
\begin{tabular}{|l| l| l|| c@{\quad}c|}
\hline \hline
$\chi(1)$ & $\chi(2)$ & $\chi(3)$ & $\mathcal{S}_{exact}$ & $1-\mathcal{H}_{exact}$ \\
\hline
\hline
$X_1$&$X_1$&$X_2$ & $\bar{p}^2 $ & $\bar{q}(1+q-q^2)$  \\
$X_1$&$X_2$&$X_1$ & $\bar{p}^2(1+p)$ & $\bar{q}(1+q-q^2)$  \\
$X_1$&$X_1X_2$&$X_1$ & $\bar{p}$ & $\bar{q}$ \\
$X_1$&$X_1X_2$&$X_2$ & $\bar{p}$ & $\bar{q}^2(1+q)$  \\
$X_1X_2$&$X_1$&$X_2$ & $\bar{p}(1+p-p^2)$ & $\bar{q}^2(1+q)$ \\
$X_1X_2$&$X_1$&$X_1$ & $\bar{p}$ & $\bar{q}$ \\
$X_1X_2$&$X_1X_2$&$X_1$ & $\bar{p}(1+p)$ & $\bar{q}^2$ \\
$X_1X_2$&$X_1$&$X_1X_2$ & $\bar{p}(1+p)$ & $\bar{q}^2$ \\
$X_1X_2$&$X_1X_2$&$X_1X_2$ & $1-p^3$ & $\bar{q}^3$ \\
\hline \hline
\end{tabular}
\caption{Different configurations and associated survivability and hackability for the simple
graph from figure\ref{fig:9Spreads}. $\mathcal{S}_{exact}$ is exact value of
survivability, $1-\mathcal{H}_{exact}$ is 1 minus exact value of hackability. } 
\label{Tablica3vertices}
\end{table}

\begin{figure}
 \includegraphics[width=0.48\textwidth]{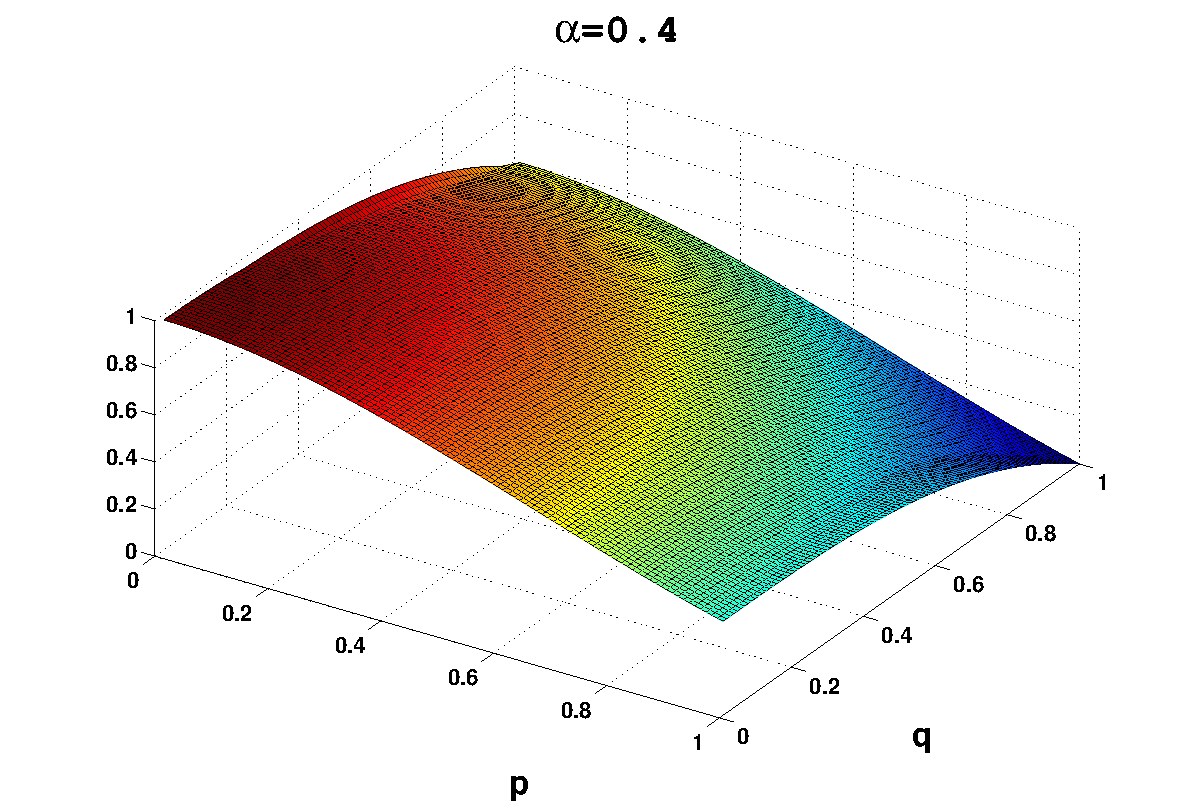}
\caption{ The plot of $\mathbf{F}\equiv max\left(\mathcal{F}(\alpha=0.4,p,q)\right)$
depending on parameters $p$ and $q$ for the choice $\alpha=0.4$.}
\label{fig:PhasePlot}
\end{figure}

\section{Examples on simple graphs}

In the following, an easy configuration problem is solved on a network
containing 3 vertices and 2 edges for all distinct configurations, as shown in Fig.\
\ref{fig:9Spreads}. There are 9 distinct configurations with respect to graph automorphism.
It is noted that no difference exists between the upper left figure and its rotation around
the $y$-axis, and no difference exists if the colors are exchanged. All these
configurations are equivalent with respect to the graph structure and the fact that all
symbols of information have essentially the same value.
For the complete graph of two vertices, 6 distinct configurations would be obtained, instead of 9.

In table~\ref{Tablica3vertices}, all exactly calculated survivabilities and hackabilities
for this simple example are listed. It can be noticed that the value of survivability is maximal in the
last row and that the value of hackability is minimal in the first and second row, as was previously concluded. In Fig.~\ref{fig:PhasePlot}, the maximal value of the
functional for different values of parameters $p$ and $q$ does not appear to show
interesting behavior; however, closer inspection reveals a discontinuity in the slope of the
maximal functional surface near extreme values of $p$.

\begin{figure}
 \includegraphics[width=0.48\textwidth]{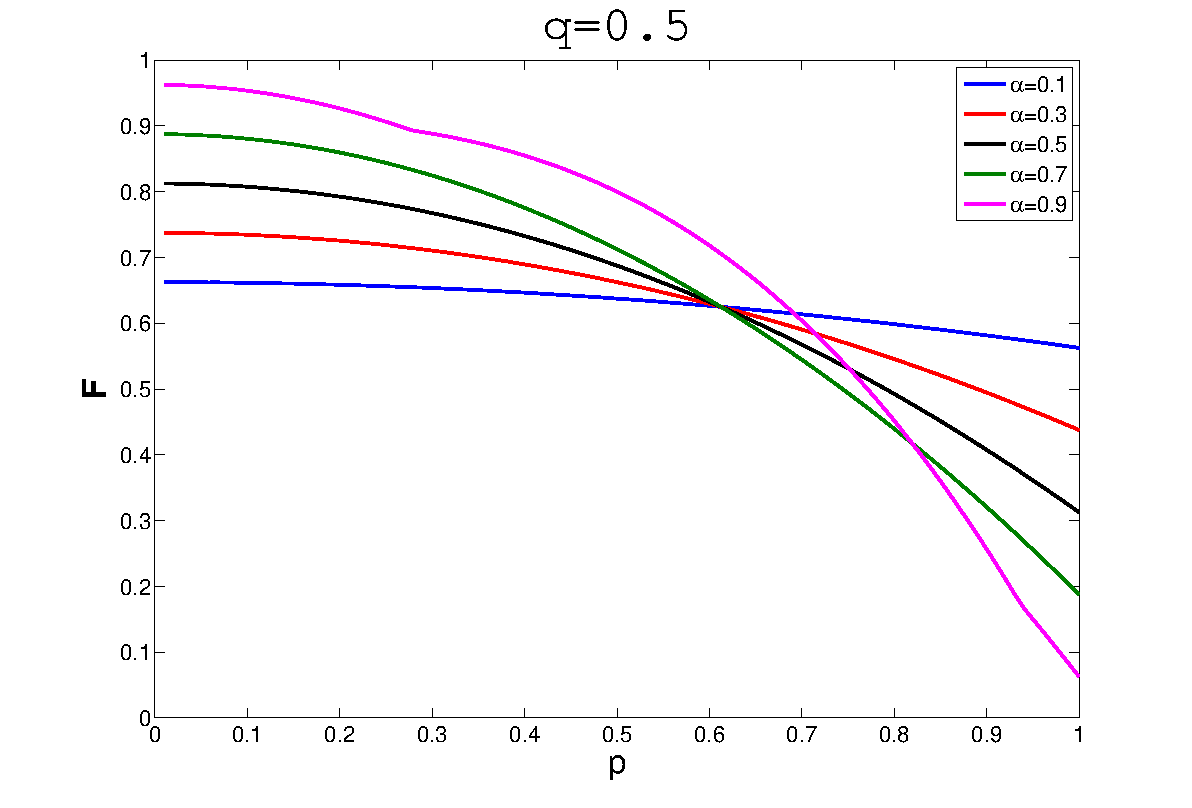}
\caption{ The plot of $\mathbf{F}\equiv max\left(\mathcal{F}(\alpha,p,q=0.5)\right)$
depending on parameters $\alpha$ and $p$ for the choice $q=0.5$.}
\label{fig:PhasePlot2}
\end{figure}

\begin{figure}
 \includegraphics[width=0.48\textwidth]{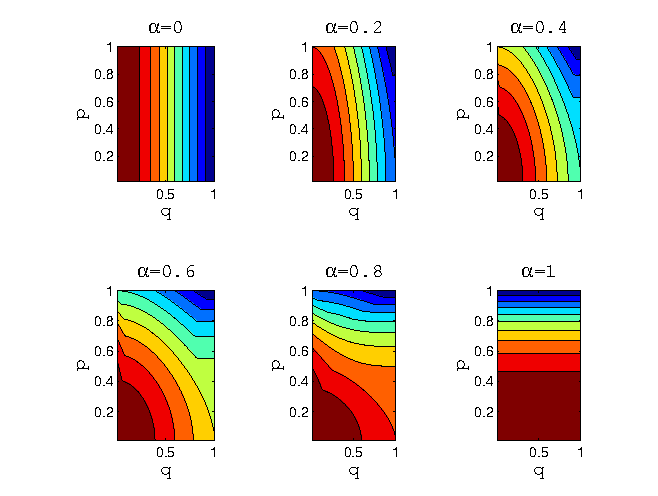}
\caption{ The surface plot of $\mathbf{F}\equiv max\left(\mathcal{F}(\alpha,p,q)\right)$
for different values of  $\alpha$, $p$ and $q$. One can notice nontrivial behavior even
in such a simple example.}
\label{fig:PhasePlot3}
\end{figure}

In Fig.~\ref{fig:PhasePlot2}, a change of regime is observed at some point in the interval $0.7<\alpha<0.9$. The set of figures
\ref{fig:PhasePlot3} indicates that different behaviors of the functional are obtained
as the parameters are changed. This relatively rich behavior is apparent for such a simple
network, and even richer behavior can be expected to be observed as the network size is increased.

\section{Derivation of Survivability and Hackability}
\label{app:derivations}

\subsection{Survivability and MICS}
First, all subgraphs of size $|\Gamma'|=1$ (i.e., single vertices) are checked for the presence of the complete information. Second, their number is counted and denoted by $n_1$. The contribution to survivability is then $n_1\pbar$. It is noted that all other parts of the network are not accounted for at this stage. If the information exists in these vertices, it will also exist in any subgraph that contains them; therefore, such larger subgraphs need not be accounted for further, because the distribution of $\pbar$ and $p$ on the remaining vertices can take arbitrary values while the corresponding subgraph still contributes to survivability. This procedure can be continued for increasingly larger subgraphs. One has to check only subgraphs $\Gamma'$ that do not contain complete
subgraph $\Gamma''$ that had complete information stored and is smaller than the subgraph $\Gamma'$ i.e. $|\Gamma'|>|\Gamma''|$. Such subgraphs will be denoted as $\tilde{\Gamma}'$ and will be named \emph{minimal
information-carrying subgraph} (MICS). It is important to stress that two different such subgraphs can have an overlap, but never can one of them be a proper subgraph of the other. While this procedure is cumbersome, finding more and more subgraphs with complete information reduces the number of larger subgraphs that have to be checked, thus making it viable for small graphs.
In the end, the inclusion-exclusion principle is used in order to exclude all double countings.
Following this procedure Survivability can be written as:

\begin{align}
\mathcal{S}(p,\chi,\Gamma)&=&\sum_{\tilde{\Gamma}'\subset\Gamma}\bar{p}^{|\tilde{\Gamma}'|}-\sum_{\tilde{\Gamma}'_i,\tilde{\Gamma}'_j}\bar{p}^{|\tilde{\Gamma}'_i\bigcup\tilde{\Gamma}'_j|}\nonumber\\
&+&\sum_{\tilde{\Gamma}'_i,\tilde{\Gamma}'_j,\tilde{\Gamma}'_k}\pbar^{|\tilde{\Gamma}'_i\bigcup\tilde{\Gamma}'_j\bigcup\tilde{\Gamma}'_k|}-\ldots 
\label{eq:Surviv-exact-sizes}
\end{align}

Notice that this equation relies on the independence of failures in the graph. In the introduction was mentioned that probabilities $p$ and $\pbar$ can also represent the probabilities of vertices being active, and in principle one could easily put joint probabilities extracted from activity data in the equation. Such more general equation is \begin{align}
\mathcal{S}(p,\chi,\Gamma)&=&\sum_{\tilde{\Gamma}'\subset\Gamma}\bar{P}(\tilde{\Gamma}')-\sum_{\tilde{\Gamma}'_i,\tilde{\Gamma}'_j\subset\Gamma}\bar{P}(\tilde{\Gamma}'_i\bigcup\tilde{\Gamma}'_j)\nonumber\\
&+&\sum_{\tilde{\Gamma}'_i,\tilde{\Gamma}'_j,\tilde{\Gamma}'_k\subset\Gamma}\bar{P}(\tilde{\Gamma}'_i\bigcup\tilde{\Gamma}'_j\bigcup\tilde{\Gamma}'_k)-\ldots 
\label{eq:Surviv-exact-sizes-joint}
\end{align} 

The terms in equation \ref{eq:Surviv-exact-sizes}, do not take care of the size of subgraphs but they can be rewritten using the size of subgraphs in the following way:

 Then $\mathcal{S}(p,\chi,\Gamma)$ is a
polynomial in $\bar p$,
\begin{equation} 
\mathcal{S}(p,\chi,\Gamma) = \sum_{r=1}^{|\Gamma|} a_r(\chi,\Gamma)\,\bar p^r , \label{eqap:SurvivabilityEXACT} 
\end{equation} 
where the coefficient of $\bar p^r$ is 
\begin{equation} a_r(\chi,\Gamma) = \sum_{\ell=1}^{M} (-1)^{\ell+1} \sum_{\substack{ 1\leq j_1<j_2<\cdots<j_\ell\leq M\\ |\tilde{\Gamma}'_{j_1}\cup\tilde{\Gamma}'_{j_2}\cup\cdots\cup \tilde{\Gamma}'_{j_\ell}|=r }} 1 . \label{eqap:SurvivabilityCoefficient} \end{equation}

{\bf Proof} 
Let $V_S\subseteq V$ denote the random set of surviving vertices. The exact survivability is the probability that the surviving network contains at least one connected vertex set carrying all symbols. Then Survivability is the probability that, after node failures, there exists some subset $U$ of the surviving nodes $V_S$ such that the subgraph induced by $U$ is connected and the nodes in $U$, taken together, contain all symbols.
 For every MICS $\widetilde{\Gamma}'\in\widetilde{\mathcal G}$, define the event 
 \begin{equation} E_{\widetilde{\Gamma}'} = \left\{ V(\widetilde{\Gamma}')\subseteq V_S \right\}. \label{appeq:survivability-event}\end{equation} 
 
 This is the event that all vertices of $\widetilde{\Gamma}'$ survive. We first show that the survivability event is exactly the union of the events
\(E_{\widetilde{\Gamma}'}\) over \(\widetilde{\Gamma}'\in\widetilde{\mathcal G}\). If $E_{\widetilde{\Gamma}'}$ occurs for some $\widetilde{\Gamma}'\in\widetilde{\mathcal G}$, then all vertices of $\widetilde{\Gamma}'$ survive. Since $\widetilde{\Gamma}'$ is connected and carries all symbols, the surviving network contains complete information. Conversely, suppose that the surviving network contains a connected vertex set $U\subseteq V_S$ such that 
 \begin{equation} 
 \bigcup_{v\in U}\chi(v)=\{1,\ldots,N\}. 
 \end{equation} 
 Because the graph is finite, among all connected subsets of $U$ carrying all symbols, there exists an inclusion-minimal one; call it $U_{\min}$. By definition, $U_{\min}$ is the vertex set of some $\widetilde{\Gamma}'\in\widetilde{\mathcal G}$. Since $U_{\min}\subseteq U\subseteq V_S$, all vertices of this minimal subgraph survive and hence $E_{\widetilde{\Gamma}'}$ occurs. Therefore 
 \begin{equation} \mathcal S(p,\chi,\Gamma) =  P \left( \bigcup_{\widetilde{\Gamma}'\in\widetilde{\mathcal G}} E_{\widetilde{\Gamma}'} \right). \label{eq:survivability-union-events} \end{equation} 
 Applying the inclusion-exclusion principle to the finite family of events $\{E_{\widetilde{\Gamma}'}\}_{\widetilde{\Gamma}'\in\widetilde{\mathcal G}}$ gives 
 \begin{equation}  P \left( \bigcup_{\widetilde{\Gamma}'\in\widetilde{\mathcal G}} E_{\widetilde{\Gamma}'} \right) = \sum_{\emptyset\neq \mathcal A\subseteq \widetilde{\mathcal G}} (-1)^{|\mathcal A|+1}  P \left( \bigcap_{\widetilde{\Gamma}'\in\mathcal A} E_{\widetilde{\Gamma}'} \right). \label{eq:inclusion-exclusion-survivability} \end{equation} 
 For a fixed nonempty subfamily $\mathcal A\subseteq\widetilde{\mathcal G}$, the intersection event 
 \begin{equation} 
 \bigcap_{\widetilde{\Gamma}'\in\mathcal A} E_{\widetilde{\Gamma}'} \end{equation} 
 is precisely the event that all vertices belonging to at least one selected minimal information-carrying subgraph survive. Hence 
 \begin{equation} 
 \bigcap_{\widetilde{\Gamma}'\in\mathcal A} E_{\widetilde{\Gamma}'} = \left\{ \bigcup_{\widetilde{\Gamma}'\in\mathcal A} V(\widetilde{\Gamma}') \subseteq V_S \right\}. 
 \end{equation} 
 Since vertex survival is independent and each vertex survives with probability $\bar p$, we have 
 \begin{equation}  P \left( \bigcap_{\widetilde{\Gamma}'\in\mathcal A} E_{\widetilde{\Gamma}'} \right) = \bar p^{ \left| \bigcup_{\widetilde{\Gamma}'\in\mathcal A} V(\widetilde{\Gamma}') \right| }. \label{eq:intersection-probability} \end{equation} 
 Substituting Eq.~\eqref{eq:intersection-probability} into Eq.~\eqref{eq:inclusion-exclusion-survivability} proves Eqs.~\eqref{eqap:SurvivabilityEXACT} and ~\eqref{eqap:SurvivabilityCoefficient}.  

This definition is closely related to the notion of a connected tropical subgraph in vertex-coloured graphs. In that literature, a subgraph is called tropical if it contains at least one vertex of every colour, and the corresponding algorithmic problem is to find connected vertex sets that realize all colours of the graph~\cite{angles2016connected,chapelle2017exact}. In the special case in which each vertex stores exactly one symbol, the symbols $X_\mu$ may be identified with colours, and an information-carrying connected subgraph is precisely a connected tropical subgraph. The MICS used here correspond to the inclusion-minimal members of this family, since no proper connected subgraph of MICS is allowed to contain the complete set of symbols. This makes the present construction particularly close to the enumeration problem for minimal tropical connected sets~\cite{kratsch2017enumerating,bliznets2023enumeration}.

There are, however, two important differences. First, in the present problem a vertex may store an arbitrary subset of symbols, so the vertex labelling is multi-valued rather than an ordinary colouring. Thus MICS are not merely tropical subgraphs of a vertex-coloured graph, but a natural multi-symbol extension of them. Second, the role of these subgraphs is different. The tropical-subgraph literature is primarily concerned with finding or enumerating small connected colourful sets, whereas here all MICS enter as elementary reconstruction events in an inclusion--exclusion expression for the survivability polynomial. 

\subsection{Hackability}
In similar way the hackability can be derived. For pedagogical reasons, the special case with only two symbols $X_1$
and $X_2$ is first analyzed. In that case the set $\Omega(\bsy{X})=\{X_1,X_2,X_1X_2\}$ is the set of all possible vertex configurations. The number of these elements in the configuration $\chi$ are given with       
$n_{X_{\mu}}=\sum_i\delta_{\chi(i),X_{\mu}}$ and
$n_{X_1X_2}=\sum_i\delta_{\chi(i),X_1X_2}$. The hackability is
\beq\label{hackabilityN2}
\mathcal{H}(q,\chi)=1-\bar{q}^{n_{X_1X_2}}(\bar{q}^{n_{X_1}}+\bar{q}^{n_{X_2}}-\bar{q}^{n_
{ X_1}+n_{X_2}}),
\eeq
where $\bar{q}=(1-q)$. It is clear that $\bar{\mathcal{H}}$ is going to be led by vertices
which have more then one symbol attached. Let us consider two configurations $\chi$ and $\chi'$
with associated number of symbols $n$ and $n'$. The first configuration contains all the possible symbol
combinations and the second maps just one symbol on every vertex. The configurations have
the same values on vertices with one symbol, but they clearly differ on vertices that
store two symbols in the first configuration. Then
$n_{X_1}+n_{X_2}+n_{X_1X_2}=n'_{X_1}+n'_{X_2}$ and also $n'_{X_{1}}=n_{X_{1}}+an_{X_1X_2}$
and $n'_{X_{2}}=n_{X_{2}}+(1-a)n_{X_1X_2}$. Comparing the 2 configurations one can show that configuration  $\chi'$ leads always to a smaller value of $\mathcal{H}$, therefore to the larger value of $\bar{\mathcal{H}}$.

 A general case of $N$ different symbols is now presented. The $\mathcal{N}(X_i)$ is defined as the number of vertices in which the symbol $X_i$ is stored  - either alone or with other symbols. The $\mathcal{N}(X_i)$ can be written as i.e.
\begin{align} \mathcal{N}(X_i)=&n_{X_1\ldots X_i\ldots X_N}+\sum_{\mu\neq i}n_{X_1\ldots\bar{X}_{\mu}\ldots X_i\ldots X_N}+\nonumber\\
&+\sum_{\mu\neq\nu\neq i}n_{X_1\ldots\bar{X}_{\mu}\ldots X_i\ldots\bar{X}_{\nu}\ldots X_N}+,\ldots,\nonumber\\
&+\sum_{\mu\neq i}n_{X_iX_{\mu}}+n_{X_i}\nonumber\\
&=\sum_{\omega(X_{\mu})}n_{\omega(X_{\mu})},
\end{align}
where $\bar{X}_{j}$ represents that j-th symbol is not counted, and $\omega(X_{\mu})\in\Omega$ represents some subset of the total set which contains symbol $X_{\mu}$, and summation in the last line is performed over all such subsets. For example in the case of just two symbols $n_{\bar{X}_1X_2}\equiv n_{X_2}$. Furthermore, it is convenient to define  operation $\circ$ as 
\begin{align}
  \mathcal{N}_{X_1}\circ \mathcal{N}_{X_2}&= 
  \left(\sum_{\omega(X_{1})}n_{\omega(X_{1})}\right)\circ\left(\sum_{\omega(X_{2})}n_{\omega(X_{2})}\right)\nonumber\\
  &=\sum_{\omega(X_{1})}n_{\omega(X_{1})}+\sum_{\omega(X_{2})}n_{\omega(X_{2})}\nonumber\\
  & -\sum_{\omega(X_1X_{2})}n_{\omega(X_1X_{2})}.\label{eqap:binary realtion}
  \end{align}
  In eq: \ref{eqap:binary realtion}, the number of all the vertices in which symbol $X_1$ is stored is added to the number of all the vertices that store symbol $X_2$ and subtracted with the number of all the vertices that store both symbols simultaneously. For the case of N=2, the equation is $\mathcal{N}_{X_1}\circ \mathcal{N}_{X_2}=n_{X_1X_2}+n_{X_1}+n_{X_2} $. The  meaning of this operation is that it counts the number of vertices that contain symbols in the argument i.e. the interpretation of $\mathcal{N}_{X_1}\circ \mathcal{N}_{X_2}$ is that the operation counts the number of all the vertices that have stored either symbol $X_1$ or the symbol $X_2$.

Then $\bar{\mathcal{H}}$ can be written as:

\begin{align}
    \bar{\mathcal{H}}&=\sum_{i}\bar{q}^{\mathcal{N}(X_i)}-\sum_{i< j}\bar{q}^{\mathcal{N}(X_i)\circ \mathcal{N}(X_j)}+\nonumber\\
    &+\sum_{i< j< k}\bar{q}^{\mathcal{N}(X_i)\circ \mathcal{N}(X_j)\circ \mathcal{N}(X_k)}+\ldots,
    \label{eqap: H_EXACT_Summation}
\end{align}
and alternating summation is continued until all the indices have been included in the last term.

\section{Semi-local robustness functional}
\label{app:semi-local-functional}

 For large graphs, the exact computation of $\mathcal S$ requires the enumeration of all inclusion-minimal connected subgraphs $\tilde{\Gamma}'\subseteq\Gamma$ whose vertices collectively contain the complete information. We therefore introduce a semi-local approximation in which only such subgraphs contained in finite-radius neighborhoods are retained. Let 
 \begin{equation} B_R(v)=\{u\in V:\operatorname{dist}(u,v)\leq R\} \end{equation} 
 be the radius-$R$ neighborhood of vertex $v$. We denote by 
 $\widetilde{\mathcal G}_v^{(R)}(\chi,\Gamma)$ 
 the family of all MICS $\tilde{\Gamma}'$ satisfying \begin{equation} \tilde{\Gamma}'\subseteq B_R(v), \qquad v\in\tilde{\Gamma}', \qquad \bigcup_{i\in\tilde{\Gamma}'}\chi(i)=\mathbf X . \end{equation} 
 
 As explained in app.\ref{app:derivations} these graphs have the additional condition that no proper connected subgraph of $\tilde{\Gamma}'$ already contains the complete information. Thus $\widetilde{\mathcal G}_v^{(R)}$ is the local family of minimal information-carrying subgraphs rooted at $v$. The probability that at least one of these local subgraphs survives is 
 
 \begin{equation} P_v^{(R)} = \sum_{\emptyset\neq \mathcal A\subseteq \widetilde{\mathcal G}_v^{(R)}} (-1)^{|\mathcal A|+1} \bar p^{ \left| \bigcup_{\tilde{\Gamma}'\in\mathcal A} \tilde{\Gamma}' \right| }. \label{eq:local-survival-tildegamma} \end{equation}

 Now there are two roads one can take depending on the computational costs, size of the network, value of $R$ etc.
 The first one is computationally easy, but can have a wrong value of Survivability even if the configuration $\chi$ is close to optimal one.  
 That is, one can approximate semi-local survivability as
 \begin{equation} \mathcal S_{R1}(p,\chi,\Gamma) = 1-\prod_{v\in V}\left(1-P_v^{(R)}\right). \label{appeq:semi-local-survivability1} \end{equation}
 
 This approximation keeps the exact inclusion--exclusion structure inside each local neighborhood, but treats the reconstruction events associated with different centers as independent. 
 
 Second route is to enumerate all the local overlaps take their statistics and pass them to the rest of the network.
 Then the semi-local survivability can be computed as follows.

For each nonempty subfamily $\mathcal A_v\subseteq \widetilde{\mathcal G}_v^{(R)}$, we define 
\begin{equation} \Gamma'_{\mathcal A_v} = \bigcup_{\tilde{\Gamma}'\in \mathcal A_v} V(\tilde{\Gamma}'), \label{appeq:gamma-prime-family} \end{equation} 
where $V(\tilde{\Gamma}')$ denotes the vertex set of the subgraph $\tilde{\Gamma}'$. Thus $\Gamma'_{\mathcal A_v}$ is the set of all vertices appearing in the selected minimal information-carrying subgraphs from the $R$-neighborhood of $v$. The semi-local survivability obtained by applying inclusion--exclusion over all neighborhood centers is then 
\begin{equation} \mathcal S_{R2}(p,\chi,\Gamma) = \sum_{\emptyset\neq \mathcal C\subseteq V} \; \sum_{\substack{ (\mathcal A_v)_{v\in\mathcal C}\\  \mathcal A_v\subseteq \widetilde{\mathcal G}_v^{(R)}  }} (-1)^{1+\sum_{v\in\mathcal C}|\mathcal A_v|} \, \bar p^{ \left| \bigcup_{v\in\mathcal C} \Gamma'_{\mathcal A_v} \right| }. \label{appeq:semi-local-survivability2} \end{equation} 
Here $\mathcal C$ is a nonempty set of neighborhood centers. For each $v\in\mathcal C$, the set $\mathcal A_v$ selects a nonempty family of minimal information-carrying subgraphs from the local neighborhood $\widetilde{\mathcal G}_v^{(R)}$. The exponent therefore counts the number of distinct vertices contained in the union of all selected local subgraphs, rather than the sum of their sizes counted separately. Here is assumed that there are no MICS in two different neighborhoods; if there are, they should be removed from all other neighborhoods except one for equation to be correct.

 The hackability term is not local in the same sense, because in the present model hacking is independent of the graph edges. It can, however, be updated from global symbol-subset counts. For every nonempty subset $T\subseteq\mathbf X$, define \begin{equation} c_T=\#\{v\in V:\chi(v)\cap T\neq\varnothing\}. \end{equation} Then \begin{equation} \bar{\mathcal H}(q,\chi) = \sum_{\emptyset\neq T\subseteq\mathbf X} (-1)^{|T|+1} \bar q^{c_T}. \label{eq:semi-local-hbar} \end{equation} The resulting semi-local functional is \begin{equation} \mathcal F_R(\alpha,p,q,\chi,\Gamma) = \alpha\mathcal S_R(p,\chi,\Gamma) + (1-\alpha)\bar{\mathcal H}(q,\chi). \label{eq:semi-local-functional} \end{equation} Thus the survivability contribution is evaluated from local minimal information-carrying subgraphs $\tilde{\Gamma}'$, while the security contribution is evaluated from symbol-frequency counters. This provides a natural compromise between exact global optimization and purely local dynamics.

\section{Analytical estimate of the approximation error}
\label{app:error-estimate}

Here we outline how the error of the local approximation to the robustness
functional can be estimated. Let $\widetilde{\mathcal G}(\chi,\Gamma)$ denote
the family of all inclusion-minimal connected subgraphs whose vertices contain
the complete information. For $C\in\widetilde{\mathcal G}$, let $E_C$ be the
event that all vertices of $C$ survive. The exact survivability is
\begin{equation}
\mathcal S
=
P\left(
\bigcup_{C\in\widetilde{\mathcal G}} E_C
\right).
\end{equation}
Suppose that the approximation keeps only a subfamily
$\widetilde{\mathcal G}_R\subseteq\widetilde{\mathcal G}$, for example those
minimal information-carrying subgraphs contained in radius-$R$ neighborhoods.
Define
\begin{equation}
\mathcal S_R^{\cup}
=
P\left(
\bigcup_{C\in\widetilde{\mathcal G}_R} E_C
\right).
\end{equation}
Since omitted covers can only increase survivability, one has
\begin{equation}
0\leq
\mathcal S-\mathcal S_R^{\cup}
\leq
\sum_{C\in\widetilde{\mathcal G}\setminus\widetilde{\mathcal G}_R}
\pbar^{|C|}.
\label{eq:omitted-cover-error}
\end{equation}
If the approximation is defined by a maximal retained cover size $M$, and
$N_m$ denotes the number of omitted minimal covers of size $m$, then
\begin{equation}
0\leq
\mathcal S-\mathcal S_M
\leq
\sum_{m>M}N_m\pbar^m .
\label{eq:size-tail-error}
\end{equation}
For a locally tree-like graph with average degree $z$, the rough estimate
$N_m\lesssim |\Gamma|z^{m-1}$ gives
\begin{equation}
\mathcal S-\mathcal S_M
\lesssim
|\Gamma|\,\pbar\,
\frac{(z\pbar)^M}{1-z\pbar},
\qquad z\pbar<1 .
\label{apeq:tree-tail-error}
\end{equation}
Thus the truncation is controlled when larger connected covers are sufficiently
suppressed by their survival probability.

The second source of error comes from combining local reconstruction events.
Let $E_v^{(R)}$ be the event that the radius-$R$ ball around $v$ contains
at least one surviving connected cover, and let
$P_v^{(R)}=P(E_v^{(R)})$. The simplest local approximation is
\begin{equation}
\mathcal S_{R1}
=
1-\prod_v\left(1-P_v^{(R)}\right).
\label{eq:product-survival}
\end{equation}
This treats the local events as independent. 

For the second survivability approximation the following holds.

The semi-local approximation does not discard all nonlocal information
indiscriminately. It accounts exactly for every minimal connected
information-carrying subgraph that is contained in at least one radius-$R$
neighborhood. To make this precise, let
\begin{equation}
B_R(v)=\{u\in V(\Gamma): d_\Gamma(u,v)\leq R\}
\end{equation}
be the radius-$R$ neighborhood of vertex $v$, and let
$\mathcal V(\widetilde{\Gamma}')$ denote the vertex set of a MICS $\widetilde{\Gamma}'$. We define
\begin{equation}
\rho(\widetilde{\Gamma}')
=
\min_{v\in V(\Gamma)}
\max_{u\in \mathcal V(\widetilde{\Gamma}')}
d_\Gamma(u,v).
\end{equation}
Thus $\rho(\widetilde{\Gamma}')$ is the smallest radius of a graph ball
containing $\widetilde{\Gamma}'$. The semi-local approximation includes exactly
those minimal subgraphs for which
\begin{equation}
\rho(\widetilde{\Gamma}')\leq R.
\end{equation}
Consequently, the only subgraphs that can contribute to the error are those
whose spatial extent is larger than the neighborhoods used in the approximation.
 
 Therefore,
\begin{equation}
0\leq S_{\rm true}-S_{R2}
\leq
\sum_{\widetilde{\Gamma}'\in\mathcal M_R}
\bar p^{\nu(\widetilde{\Gamma}')},
\qquad
\nu(\widetilde{\Gamma}')
=
|\mathcal V(\widetilde{\Gamma}')|.
\end{equation}
Equivalently,
\begin{equation}
0\leq S_{\rm true}-S_{R2}
\leq
\sum_{\substack{
\widetilde{\Gamma}'\in\widetilde{\mathcal G}\\
\rho(\widetilde{\Gamma}')>R
}}
\bar p^{\nu(\widetilde{\Gamma}')}.
\end{equation}
In particular, if all globally minimal connected information-carrying subgraphs
are contained in some radius-$R$ neighborhood, then
\begin{equation}
S_{R2}=S_{\rm true}.
\end{equation}

For hackability, if all symbol subsets are included, the expression for
$\bar{\mathcal H}=1-\mathcal H$ is exact:
\begin{equation}
\bar{\mathcal H}
=
\sum_{\emptyset\neq T\subseteq \mathbf X}
(-1)^{|T|+1}
\qbar^{c_T},
\qquad
c_T=\#\{v:\chi(v)\cap T\neq\emptyset\}.
\end{equation}
If this inclusion--exclusion sum is truncated at order $K$, Bonferroni bounds
give a rigorous error estimate. Writing
\begin{equation}
B_K=
\sum_{1\leq |T|\leq K}
(-1)^{|T|+1}\qbar^{c_T},
\end{equation}
one has
\begin{equation}
B_{2m}\leq \bar{\mathcal H}\leq B_{2m-1}.
\label{eq:bonferroni-hbar}
\end{equation}
Therefore, if the midpoint
$\bar{\mathcal H}_{2m}^{\rm mid}=(B_{2m}+B_{2m-1})/2$ is used, then
\begin{equation}
\left|
\bar{\mathcal H}
-
\bar{\mathcal H}_{2m}^{\rm mid}
\right|
\leq
\frac{B_{2m-1}-B_{2m}}{2}.
\end{equation}

It should be noted that for practical purposes $S_{R2}$ can also be truncated, and the truncation error analyzed through Bonferroni bounds.

Combining the above estimates, the error of the approximate robustness
functional
\begin{equation}
\mathcal F_R=\alpha\mathcal S_R+(1-\alpha)\bar{\mathcal H}_R
\end{equation}
is bounded schematically by
\begin{equation}
|\mathcal F-\mathcal F_R|
\leq
\alpha\epsilon_{\rm cover}+ (1-\alpha)\epsilon_{\rm hack},
\label{appeq:functional-error-bound}
\end{equation}

where $\epsilon_{\rm cover}$ is estimated by Eq.~\eqref{eq:size-tail-error}, and $\epsilon_{\rm hack}$ by the Bonferroni gap when the hackability expansion is truncated.

\section{Semi-local message-passing optimization}
\label{app:semi-local-message-passing}

The exact optimization of the robustness functional is difficult because the
survivability term depends on all minimal information-carrying subgraphs and
on their overlaps. For this reason we used a semi-local max-sum procedure as a
heuristic optimization method. The aim of the algorithm is not to evaluate the
exact global functional during the optimization, but to generate good candidate
placements from local information.

Let $\Gamma=(V,E)$ and let
$\bsy X={X_1,\ldots,X_{N_{\rm sym}}}$ be the set of symbols. The state of
node $v$ is a non-empty binary vector
\begin{equation}
\chi_v\in\Omega_v\subseteq {0,1}^{N_{\rm sym}}\setminus{0},
\end{equation}
where $\chi_{v,a}=1$ means that $v$ stores symbol $X_a$. In the numerical
implementation, $\Omega_v$ is restricted to states containing at most $M$
symbols per node. Leaves are further restricted to singleton states, since
allowing them to store many symbols greatly increases the local state space
without producing many independent reconstruction paths.

For each node $c$ we define a local factor with scope
\begin{equation}
B_c={c}\cup N_c^{\rm sel},
\end{equation}
where $N_c^{\rm sel}$ is a selected subset of neighbors. The size of $B_c$ is
capped in order to keep the number of local assignments manageable. For each
assignment $\chi_{B_c}$, all connected subsets
$C\subseteq B_c$ that contain all symbols and are minimal with this property
are enumerated. A cluster of size $|C|$ survives with probability
$\pbar^{|C|}$, where $\pbar=1-p$. We assign it the additive weight
\begin{equation}
w(C)=-\log\left(1-\pbar^{|C|}\right).
\end{equation}
If the same cluster is visible in several factor scopes, its weight is divided
by its visibility multiplicity $m(C)$. The local survivability score is thus
\begin{equation}
W_c(\chi_{B_c})
=
\sum_{C\in\mathcal C_c(\chi_{B_c})}
\frac{w(C)}{m(C)} ,
\end{equation}
where $\mathcal C_c(\chi_{B_c})$ is the set of local minimal
information-carrying clusters visible from factor $c$.

The competing local term is the probability that the compromised nodes inside
$B_c$ contain all symbols. It is computed by inclusion--exclusion,
\begin{equation}
H_c(\chi_{B_c})
=
\sum_{A\subseteq \bsy X}
(-1)^{|A|}
(1-q)^{n_c(A)},
\end{equation}
where $n_c(A)$ is the number of nodes in $B_c$ carrying at least one symbol
from $A$, and $n_c(\emptyset)=0$. The local factor used in max-sum is then
\begin{equation}
\psi_c(\chi_{B_c})
=
 \alpha W_c(\chi_{B_c})+
(1-\alpha)(1-H_c(\chi_{B_c})) .
\end{equation}

Messages are sent from factors to variables. With
$m_{c\to v}^{(t)}(\chi_v)$ denoting the message from factor $c$ to node $v$,
the update is
\begin{equation}
\widehat m_{c\to v}^{(t+1)}(\chi_v)
=\max_{\chi_{B_c\setminus v}}
\left[
\psi_c(\chi_{B_c})
+
\sum_{\substack{u\in B_c\setminus v \\ d:\ u\in B_d\ d\neq c}}
m_{d\to u}^{(t)}(\chi_u)
\right].
\end{equation}
The messages are normalized by subtracting their mean and damped according to
\begin{equation}
m_{c\to v}^{(t+1)}
=\lambda m_{c\to v}^{(t)}
+
(1-\lambda)\widehat C_{c\to v}^{(t+1)},
\end{equation}
where $0\leq \lambda \leq 1$ is the damping parameter and the constant
$C_{c\to v}^{(t+1)}$ is chosen so that
\begin{equation}
\sum_{\chi_v\in\Omega_v}
m_{c\to v}^{(t+1)}(\chi_v)
=0 .
\end{equation}
Equivalently, $C_{c\to v}^{(t+1)}$ is the mean of the damped message over all
allowed states $\chi_v\in\Omega_v$. The damping reduces oscillations of the
loopy max-sum iteration, while the normalization fixes the arbitrary additive
gauge of the messages. After convergence, the state of each node is decoded
by maximizing the sum of incoming messages.

The resulting placement is then evaluated globally. All local minimal clusters
found in the neighborhoods are collected, duplicates are removed, and clusters
containing smaller complete-information clusters are discarded. If
$\mathcal C_{\rm MP}$ is the resulting set, the approximate survivability is
\begin{equation}
S_{\rm MP}
=
1-
\prod_{C\in\mathcal C_{\rm MP}}
\left(1-\pbar^{|C|}\right).
\end{equation}
This expression is exact for disjoint clusters and neglects higher-order
overlap corrections otherwise. The hackability is evaluated globally as
\begin{equation}
H_{\rm MP}
=
\sum_{A\subseteq\bsy X}
(-1)^{|A|}
(1-q)^{|U(A)|},
\end{equation}
where $U(A)$ is the set of nodes carrying at least one symbol from
$A\subseteq\bsy X$. The final approximate objective is
\begin{equation}
F_{\rm MP}
=
\alpha S_{\rm MP}
+
(1-\alpha)(1-H_{\rm MP}) .
\end{equation}

\begin{figure}
 \includegraphics[width=0.5\textwidth]{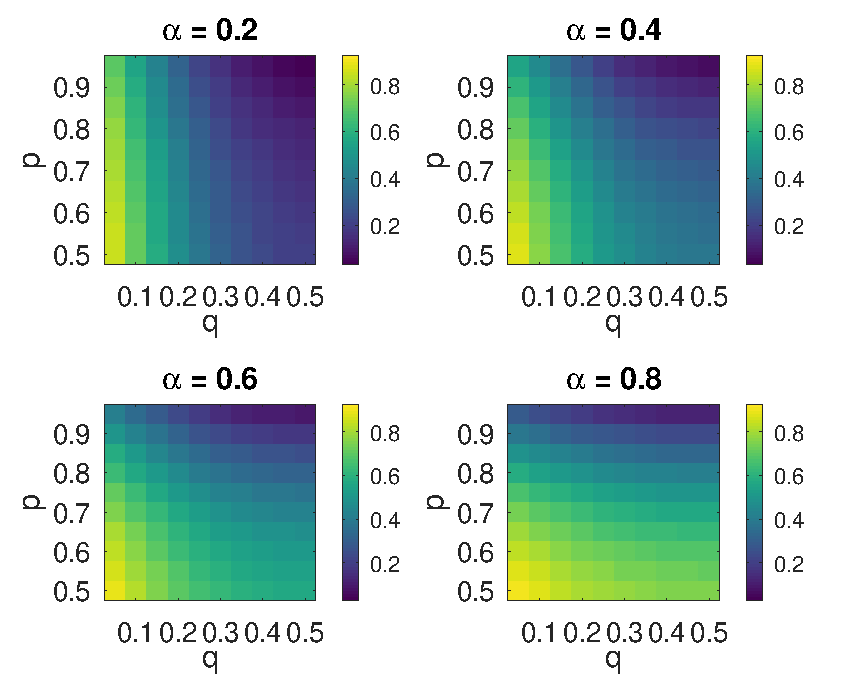}
\caption{ The plot of $\mathbf{F}\equiv max\left(\mathcal{F}(\alpha\in\{0.2,0.4,0.6,0.8\},p,q)\right)$
depending on parameters $p$ and $q$, for ER network of size 50 and $z=3$.}
\label{fig:MessagePassing}
\end{figure}

The main computational bottleneck is the construction of the factor tables.
If $s_v=|\Omega_v|$, the number of assignments inspected by factor $c$ is
\begin{equation}
A_c=\prod_{v\in B_c}s_v .
\end{equation}
Thus the method scales exponentially with the factor size and with the number
of allowed symbols per node. For this reason the scopes $B_c$ are capped and
the state spaces $\Omega_v$ are pruned. A rough cost estimate is
\begin{equation}
T_{\rm MP}
=
O\left(
\sum_{c\in V} A_c2^{|B_c|}
+
I\sum_{c\in V}A_c|B_c|
\right),
\end{equation}
where the first term corresponds to factor construction and the second to
$I$ max-sum iterations. Consequently, the method is useful for moderate
networks and small symbol sets, but it should be regarded as a heuristic
semi-local optimizer rather than a scalable exact algorithm. In figure \ref{fig:MessagePassing}, a plot of $N=50$ ER network with $z=3$ and $4$ symbols was computed, for different values of $\alpha$, $p$ and $q$. 

\section{Spin System}
\label{app:spin}

For the case of the two symbols, it is easy to verify that the equation for robustness is 

\begin{align}
 \mathcal{F}=&\alpha\left(1-p^{n_{X_1X_2}}\left(1-\pbar^2\right)^{l_{X_1X_2}}\right)\nonumber\\
 &+(1-\alpha)\left(\qbar^{\Ncal_{X_1}}+\qbar^{\Ncal_{X_2}}-\qbar^N\right).
\end{align}

Here $l_{X_1X_2}$ represents the number of edges that contain the symbol $X_1$ on one of its vertices and the symbol $X_2$ on the other. 

It is also instructive to check the expression for robustness for small values of $\pbar$ and small values of $q$. In that case, it is easy to show that the equation has the form 

\beq
 \mathcal{F} \simeq 1-\alpha+n_{X_1X_2}\left(\alpha \pbar-(1-\alpha)q\right)+\alpha\pbar^2\left(l_{X_1X_2}-\binom{n_{X_1X_2}}{2}\right).
 \label{1dspin}
\eeq

A nice property of this equation is that it is again completely local to evaluate, and it can be interpreted as a negative of the Hamiltonian of a certain type of antiferro spin system. 

Equation~\eqref{1dspin} can therefore be written as
\begin{equation}
\mathcal{F}
\simeq
1-\alpha
+
hn_{X_1X_2}
+
J
\left(l_{X_1X_2}-\binom{n_{X_1X_2}}{2}\right)
\label{appeq:spin-one-functional}
\end{equation}
where
\begin{equation}
h=\alpha\pbar-(1-\alpha)q,
\qquad
J=\alpha\pbar^2 .
\end{equation}
Equivalently, defining an effective energy
$\mathcal{E}=-\mathcal{F}$ and dropping the additive constant, we obtain
\begin{equation}
\mathcal{E}
=
-hn_{X_1X_2}
-
J
\left(l_{X_1X_2}-\binom{n_{X_1X_2}}{2}\right).
\label{appeq:spin-one-hamiltonian}
\end{equation}
The first term is an on-site field controlling the energetic cost or benefit of
placing the full information on one vertex. The second term is an
antiferromagnetic-like interaction between the pure states $X_1$ and $X_2$:
it rewards neighboring vertices that carry complementary symbols. The sign of
\begin{equation}
h=\alpha\pbar-(1-\alpha)q
\end{equation}
separates two regimes. For $h>0$, survivability dominates and vertices
carrying $X_1X_2$ are favored. For $h<0$, hackability dominates and such
vertices are penalized, while complementary neighboring single-symbol vertices
are still rewarded through the coupling $J>0$. The third term corresponds to self-field of vertices with $\chi=X_1X_2$.

This model has interesting ground state for a tradional 1d chain. Let
\begin{equation}
1\equiv X_1,\qquad 2\equiv X_2,\qquad 3\equiv X_1X_2.
\end{equation}

Apart from the constant $1-\alpha$, the functional to be maximized is then
\begin{equation}
\Phi = h n_3 + J l_{12}-J\binom{n_3}{2},
\label{eq:spin-functional}
\end{equation}
where $n_3=n_{X_1X_2}$ and $l_{12}=l_{X_1X_2}$. On an open chain of
$N=|\Gamma|$ vertices, fixing $n_3=k$ immediately determines the optimal
arrangement: the $k$ double-symbol vertices should be grouped at one boundary,
while the remaining vertices alternate between $X_1$ and $X_2$. Thus the optimal
configuration in the sector $k$ is of the form
\begin{equation}
\underbrace{33\cdots 3}_{k}\,1212\cdots ,
\end{equation}
up to reflection and exchange of $X_1$ and $X_2$. For $0\le k\le N-1$ this gives
\begin{equation}
l_{12}^{\max}(k)=N-1-k,
\end{equation}
and hence
\begin{equation}
\Phi_k
=
J(N-1)+kh-\frac{J}{2}k(k+1).
\label{eq:chain-sector}
\end{equation}
The difference between two neighbouring sectors is
\begin{equation}
\Phi_{k+1}-\Phi_k=h-J(k+1).
\end{equation}
Therefore the unconstrained optimum is obtained at
\begin{equation}
k_{\rm unc}=\left\lfloor \frac{h}{J}\right\rfloor
=
\left\lfloor
\frac{\alpha\bar p-(1-\alpha)q}{\alpha\bar p^2}
\right\rfloor ,
\end{equation}
with the usual degeneracy at integer values of $h/J$.

However, Eq.~\eqref{appeq:spin-one-hamiltonian} is itself an asymptotic expansion and is valid
only as long as
\begin{equation}
k\bar p=n_{X_1X_2}\bar p\ll 1
\end{equation}
(and similarly $kq\ll 1$ for the linear hackability expansion). Thus the
controlled ground state of the truncated theory is the boundary-defect state
\begin{equation}
\underbrace{33\cdots 3}_{k_*}\,1212\cdots
\end{equation}

with
\begin{equation}
k_*=
\min\left\{
N-1,\,
\left\lfloor\frac{\varepsilon}{\bar p}\right\rfloor,\,
\left\lfloor
\frac{\alpha\bar p-(1-\alpha)q}{\alpha\bar p^2}
\right\rfloor
\right\},
\qquad \varepsilon\ll 1 .
\end{equation}
This solution shows that, within the controlled expansion, the corrected
second-order term suppresses an extensive condensation of $X_1X_2$ vertices.
Instead, the chain consists of an alternating $X_1/X_2$ background decorated by
a finite boundary block of double-symbol vertices even if the chain is infinite.

The spin-one form of Eq.~\eqref{appeq:spin-one-functional} also clarifies the
connection with antiferromagnetic frustration. For a triangle and two symbols,
let $3\equiv X_1X_2$ denote a vertex carrying the complete information, and let
$1\equiv X_1$, $2\equiv X_2$ denote pure-symbol vertices. In the regime
$q\ll \pbar\ll 1$, keeping terms up to order $\pbar^2$, the relevant
truncated functional can be written as
\begin{align}
\mathcal F
\simeq &
1-\alpha
+
h n_3
+
J\left(l_{12}-\binom{n_3}{2}\right),
\\
h=& \alpha\pbar-(1-\alpha)q,
\\
J=&\alpha\pbar^2 .
\label{eq:triangle-truncated-functional}
\end{align}
 The term $J l_{12}$ is
antiferromagnetic-like: it rewards neighboring pure vertices carrying different
symbols. On a triangle, this interaction is frustrated, because three pure
vertices cannot make all three edges of type $1-2$. The best pure state is
therefore $112$ or $122$, with two satisfied $1-2$ edges and one
unsatisfied equal-symbol edge.

The competing candidate states have
\begin{equation}
\begin{array}{c|c|c|c}
\text{state} & n_3 & l_{12} &
\mathcal F-(1-\alpha) \\ \hline
112 \text{ or } 122 & 0 & 2 & 2J \\
312 & 1 & 1 & h+J \\
333 & 3 & 0 & 3h-3J .
\end{array}
\end{equation}
Thus the frustrated pure-symbol state is optimal for $h<J$, the mixed state
with one $C$ vertex and an $A-B$ edge is optimal for
$J<h<2J$, and the fully redundant state $CCC$ is optimal for $h>2J$. In
terms of the original parameters, the two boundaries are
\begin{equation}
\alpha_1
=
\frac{q}{q+\pbar-\pbar^2},
\qquad
\alpha_2
=
\frac{q}{q+\pbar-2\pbar^2}.
\label{eq:triangle-alpha-boundaries}
\end{equation}
Therefore,
\begin{equation}
\begin{cases}
112 \text{ or } 122, & \alpha<\alpha_1,\\
312, & \alpha_1<\alpha<\alpha_2,\\[2mm]
333, & \alpha>\alpha_2 .
\end{cases}
\end{equation}
These boundaries are consistent with the expansion when
$q\ll\pbar\ll1$. In this regime $\alpha_1,\alpha_2=O(q/\pbar)\ll1$, while
the width of the intermediate mixed phase is $O(q)$. Although the transition
is produced by a cancellation in the field
$h=\alpha\pbar-(1-\alpha)q$, the retained coupling scale is
$J=O(q\pbar)$, whereas the neglected hackability terms are $O(q^2)$ and the
neglected survivability terms are $O(q\pbar^2)$. Both are smaller under the
ordering $q\ll\pbar\ll1$. Thus the intermediate phase is narrow but remains
within the controlled Hamiltonian regime.
Therefore, surprisingly, there exists a parameter range for which frustration beats the double symbol relaxation.

\newpage

\end{document}